# Laser Beam Shaping Using a Photoinduced Azopolymer Droplet-Based Mask

[1]R. Barille, [2]A. Korbut, [2]S. Zielinska, [2]E. Ortyl, [3]D. G. Perez

[1]Univ Angers, CNRS, MOLTECH-ANJOU, SFR MATRIX, F-49000 Angers, (France)
[2]Department of Polymer Engineering and Technology, Faculty of Chemistry,
Wroclaw University of Technology, 50-370 Wroclaw, (Poland)
[3]Instituto de Física, Pontificia Universidad Católica de Valparaíso (PUCV), 23-40025 Valparaíso, (Chile)

regis.barille@univ-angers.fr

The dewetting of an azopolymer droplet, followed by the photostructuration of the evaporated droplet, is employed to create an amplitude mask. This straightforward process yields a large area featuring periodic micro- and nanostructures. The resulting pattern is utilized to generate a non-diffracting beam. Starting with a Gaussian beam illuminating the amplitude mask, the critical aspect is the production of a bright, ring-shaped beam with a high radius-to-width ratio and symmetric central laser spots, each with the same intensity. This alternative approach to shaping a laser beam is demonstrated as a rapid and cost-effective fabrication technique.

# Laser Beam Shaping Using a Photoinduced Azopolymer Droplet-Based Mask

1. Introduction

Recently, structured light beams and in particular non-diffracting beams have received a lot of attention due to their interesting properties and their applications in both fundamental and applied physics. Many studies on non-diffracting beams in micro-imaging [1] have also drawn considerable attention in fields such as free-space [2] or quantum communication [3]. In applications within the atmospheric environment, such as optical communication, non-diffracting beams demonstrate greater resilience against perturbations when compare to Gaussian beams [4].

For example, Bessel beams are an important member of the family of non-diffracting beams [5] as they are not only the most commonly used but also the first historically. Various methods have been proposed to generate this beam shape [6]. Recently a theoretical and experimental study explored both the impact of ring dimensions on the quality of the generated Bessel beam as well as its output power [7]. Furthermore, additional researches has been conducted with more complex and distinct beams. In this context, autofocusing ring Airy beams have shown a significant interest due to their ability to dramatically increase the intensity at the focal point by several orders of magnitude [8]. Autofocusing is achieved through the effective surface tension of the narrow annular area that carries the optical power. This property can be particularly useful for focusing the beam on a selected target. Characteristics of the abrupt autofocusing, including focal intensity and position, as well as subsequent defocusing can be controlled by the set-up's parameters [9].

Since the experimentation of the Airy beam [10], various other types of Airy beams have drawn considerable attention. For instance, a range of Airy beam distortions such as the ring Airy beam (RAB) [11], the Airy vortex beam [12] and the symmetric Airy beam [13] have been reported with potential applications in optical communication. Efremidis et al. provide a summary of the properties of the Airy beam [14]. Recently, radially symmetric ring Airy Gaussian vortex (RAiGV) beams have attracted a great interest due to their unique characteristics of self-focusing and self-healing [15]. These beams rapidly concentrate their energy at the focal point while maintaining a low intensity profile before this point. The maximum laser intensity experiences a dramatic increase by several orders of magnitude precisely at the focal point.

For the generation of these non-diffracting beams, various techniques are employed to transform a Gaussian beam into a structured light. One of these techniques involved the use of holograms [16]. Structured light approaches initially relied solely on amplitude control with transmissive masks, and later phase controls was achieved using computer generated holograms (CGHs). The introduction of spatial light modulators (SLMs) enables the adjustment of the local refractive index of each pixel, $n(x, y)$. This has resulted in an increase of structured light techniques using SLMs [17] contributing to the development of applications, particularly in optical communication, imaging microscopy laser material processing and optical trapping or tweezing.

In this study, an alternative approach for generating an unconventional non-diffracting beam is demonstrated. The photonic device is based on a pattern created through the dewetting of a photochromic material droplet, which can later be photoinduced for further structuring. Patterns formed by polymer molecules are of interest when it comes to diffracting beams. The material used in this study is an azopolymer, in which the photoisomerization reaction of azobenzene molecules can induce material motions at the molecular level, resulting in macroscopic changes at the surface on a larger scale [18]. In particular, amorphous azopolymer films can create stable surface relief patterns when exposed to light. This material allows for the straightforward generation of large-area periodic

micro- and nanostructures. The amorphous azopolymer used in this study enables efficient photoinduced surface-pattern formation, believed to be driven by continuous trans-cis-trans cycling of the azobenzene molecules.

Beginning with a Gaussian beam, the critical aspect is to generate a bright, ring-shaped beam with a large radius-to-width ratio, central spots within it and autofocusing properties. The method for creating this pattern involves two steps: i) controlling droplet evaporation, and ii) inducing structural changes through photosensitivity. The method presented here is flexible, easily achievable using the described material, doesn't necessitate any special optical elements, and is, therefore, applicable in various experiments. Simulations have been conducted to illustrate the beam's behaviour during its propagation.

2. Methods

2.1 Azopolymer preparation

The azopolymer is made from a highly photoactive azobenzene derivative containing heterocyclic sulfonamide moieties, which has shown its huge capacity for surface patterning. The details of synthesis of 3-[4-[(E) - (4-[(2, 6-dimethyl-pyrimidin-4-yl)amino]sulfonylphenyl)diazenyl] phenyl-(methyl)amino]propyl2-methylacrylate are reported elsewhere [19]. The solution was prepared by dissolving the azopolymer powder in THF (50 mg/mL THF). The solution was filtered through a 0.45 μm membrane. The molecular mass of the used azopolymer determined by GPC (gel permeation chromatography) was in the range of 14 000 and 19 000 g/mol. The glass transition temperature ($T_g$) is 57 °C. The absorbance at the working wavelength of 473 nm is 1.6 for a thin film with a thickness of 950 nm on a glass substrate spin-coated with the azopolymer solution.

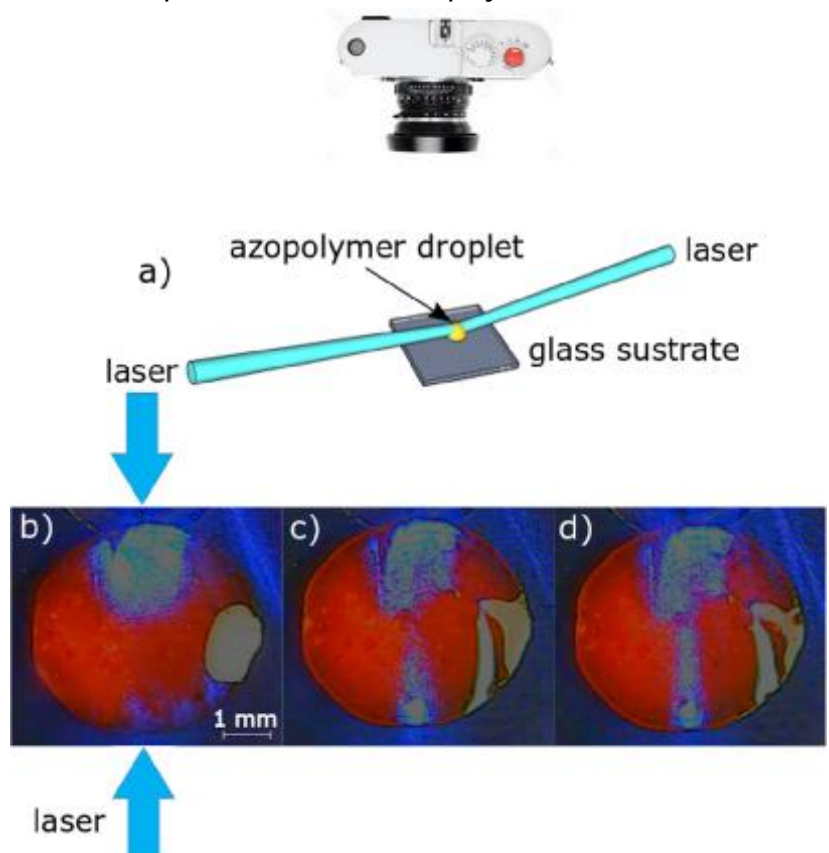

Fig. 1. a) Set-up for illumination of the droplet, b) the two counter-propagating laser beams (blue) inside the azopolymer droplet (orange) at half the apex.

## 2.2 Droplet dewetting

The pattern on the glass substrate surface is obtained using the droplet dewetting technique. A 6 µl droplet of the azopolymer solution is dispensed onto the surface of a glass substrate using a calibrated pipette. Two counter-propagating laser beams (each with a power of 100 mW) are generated by splitting a laser beam (CNI CW laser @473 nm, 400 mW) with a beam splitter and then attenuated using a neutral filter. These laser beams are directed toward the droplet near the glass surface and at the contact line of the droplet (see Figure 1a). The two beams are directed into the center of the droplet with a slight angle (5°), and they propagate inside the droplet at about half of its apex. As the droplet evaporates, these two laser beams induce thermal effects within it, specifically thermal Bénard-Marangoni convection [20]. The duration of droplet evaporation is controlled by the laser and is terminated once the solvent has completely evaporated, leaving a dried pattern. The timing is determined by observing the transition of the droplet from a liquid volume to a dried surface pattern. For a 6 µl droplet, complete drying takes approximately 20 seconds or less when using two 100 mW laser beams. In contrast, without a laser source, evaporation would take an estimated 8 to 9 minutes at ambient temperature. Figures 1b, 1c, and 1d illustrate the evolution of the droplet as the two beams heat the volume. As the volume decreases, the two blue beams become clearly visible on the top of the azopolymer droplet.

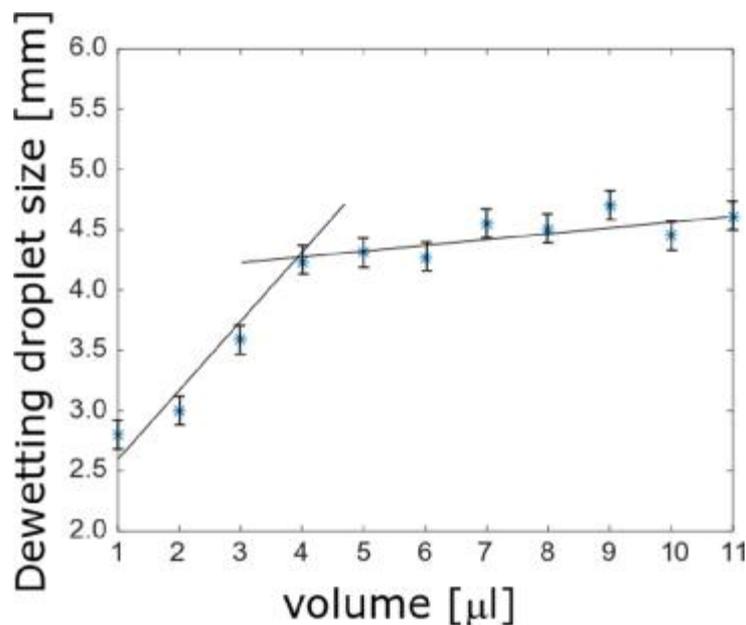

Fig. 2. Evolution of the pattern size of a dewetting droplet on a glass substrate as a function of the initial droplet volume.

Figure 2 depicts the diameters of dewetted droplet patterns on the surface of the glass substrate as a function of the initial droplet volume. Each data point represents the average value of three evaporated droplet diameters, with an error margin of 0.3 mm taken into account in their estimation.

The dewetted pattern diameter increases with the droplet volume, following a linear trend until it reaches 4 µl, at which point there is a change in slope. Beyond 4 µl, the slope becomes very small, and it appears to approach an asymptote. As anticipated, the relationship between the volume of the evaporating droplet and time is not linear, as discussed in references [21, 22].

This observation can be attributed to the viscosity of the azopolymer droplet and the rapid evaporation of the solvent, which constrains the expansion of the dried surface. The viscosity of the droplet is primarily governed by the viscosity of diluted PMMA in THF (50 mg/ml). Consequently, larger droplet volumes entail a greater amount of solvent that needs to evaporate, leading to limitations in the surface expansion of the droplet.

The spreading process can be divided into two stages:

i) An initial, brief stage where evaporation can be disregarded, and the droplet spreads with an approximately constant volume.

ii) A subsequent, slower stage in which the spreading process is nearly complete, and the evolution is primarily determined by evaporation [23].

The results indicate that evaporation can never be ignored, and for droplet volumes exceeding 6 μl, the second stage becomes predominant, ultimately constraining the droplet's spread.

Figure 3 provides details of the pattern following droplet dewetting with solvent evaporation, followed by photostructuring. The images (Fig. 3) presented here were captured using a high-resolution numerical microscope (Keyence VHX-7000). Figure 3(a) was acquired in transmission mode with an axial beam. The dewetted droplet is characterized by a large, uniformly thin ring with a diameter of 4.1 mm ± 0.2 mm and a width of 487 μm ± 10 μm, surrounding a significant central region with a diameter of 1.25 mm ± 0.2 mm.

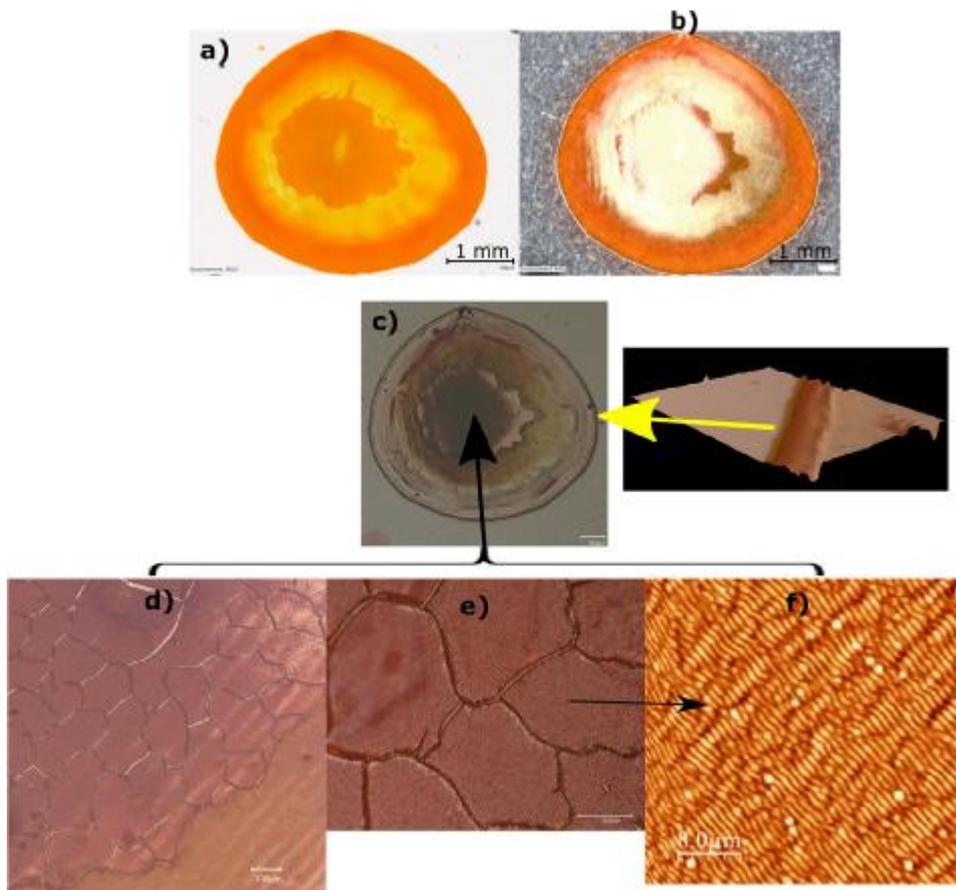

Fig. 3. Photoinduced pattern of the dewetted azopolymer droplet. The centre part of the surface structure is illuminated by a laser beam to produce a surface relief grating with a) dewetting pattern observed in transmission, b) dewetting pattern observed with an annular contrast enhancement with the ring structure topography, c) dewetting pattern observed in reflection, d) cracks observed inside

the central part of the pattern are detailed in e) and f) with photoinduced self-structures giving a surface relief grating.

The central part constitutes 50% of the total dewetted surface pattern area. Figure 3(c) illustrates the sample illuminated by a combination of an axial and annular beam, while Fig. 3(b) provides an image of the dewetted droplet in reflection with a partially annular illuminating beam for reference. In Fig. 3(c), various structures within the pattern are visible. The round central part inside the pattern exhibits a wavy shape with peaks. The distance between the peaks measures approximately 367 μm ± 50 μm and is dependent on the initial volume of the droplet. The peaks in the central part diminish as the droplet volume increases. The wavy shape is a consequence of the contact lines of the droplet volume, which initially remained pinned at the beginning of the evaporation process and are unevenly altered by evaporation. During the drying process, rings form when the contact lines remain pinned. After a certain time, the entire contact line begins to recede and move, but not uniformly. This phenomenon is referred to as the 'coffee-ring' effect, reminiscent of the familiar stains left by dried coffee drops (Fig. 3(c)) [24]. The two ring structures are a result of the higher concentration or the high molecular weight of the azopolymer solution. The entanglement of polymer chains is enhanced, and the collective behavior of polymer molecules predominates [25]. The evaporation rate is highest near the droplet's periphery due to the large contact area created by the liquid's curvature. It is further accelerated by the laser beam in proximity to the contact line between the droplet and the substrate. A Marangoni flow occurs as a result of the surface tension gradient on the free surface of the droplet caused by the non-isothermal interface, leading to efficient convective mixing within the droplet. As the solvent rapidly evaporates, the contact line quickly moves toward the center of the droplet due to the Marangoni flow, leaving behind the azopolymer molecules [26]. Two steps are observed, one at the beginning of the evaporation process leading to the formation of the outer coffee ring and a second one in the central part of the structure. It's worth noting that the central part has a higher concentration of azopolymer, confirming that this second process is slower than the first one. In the large outer ring, the concentration of azopolymer is significantly lower, and this can be observed in the topography of the ring structure (Fig. 3(c)), with the ring's width measuring about 80 μm.

The central motivation behind this setup is its application for laser beam shaping. The goal is to induce a thermally-driven Marangoni effect within the droplet, aiming to generate a recirculating thermal flow inside the droplet. The primary objective is to counteract or reduce the coffee-ring effect [27, 28] and to expand the deposition pattern size of azopolymer solution droplets containing volatile solvents.

The expected outcome of this experiment is the domination of the Marangoni effect over the coffee-ring effect, ultimately leading to its attenuation. With a mitigated coffee ring, the dewetting patterns reveal two distinct parts: a central region and a coffee-ring structure. To achieve this, the chosen approach involves creating an environment in which the evaporation rate at the apex of the droplet significantly outpaces the evaporation rate on the surrounding surfaces.

To achieve this, two laser beams are directed at the top of the droplet at a slight angle, penetrating the droplet and concentrating the laser intensity on its upper part. The intense evaporation rate induced at the central area of the droplet serves as the driving force required for the Marangoni flow, effectively mitigating the coffee-ring effect and increasing the deposition of material within the ring structure.

The size of the droplets is similar to the beam diameter, and the laser generates a laser-induced differential evaporation, as previously observed using a $CO_2$ laser on fluorescent ssDNA droplets [29]. The coffee-ring pattern is explained by an outward capillary flow. As the droplet evaporates, the two laser beams promote a maximum evaporative flux at the droplet's top. This creates an inward and radial flow, which counters the outward radial flow associated with the coffee-ring effect. The inward flow counteracts the de-pinning of the contact line and, in the end, results in the formation of peak deposition patterns within the central part of the pattern.

Self-organized structures, in the form of cracks, are only observed within the central pattern. These crack structures are consistently spaced in the central part of the evaporated structure and have dimensions of 172 μm ± 10 μm (see Fig. 3(d) and 3(e)). These cracks result from the interplay between high thermal stresses and preparation-induced stresses generated by the two laser beams, in conjunction with film slip during the dewetting process. The rapid evaporation of the droplets leads to a rapid reduction in the solution volume, which in turn results in significant in-plane tensile stresses. These stresses are directly responsible for the formation of cracks [30, 31]. The formation of cracks is significantly reduced when the droplet is allowed to evaporate without the influence of lasers and is practically absent in such cases.

2.3. Inscription of surface relief gratings (SRG)

The incoming light intensity is regulated using various neutral filters. The polarization direction of the single Gaussian laser beam is adjusted by employing a half-wave plate. The sample is positioned perpendicular to the incident laser beam. The size of the single collimated laser beam that impinges on the azopolymer dewetted pattern is controlled using a variable beam expander (Thorlabs BE052-A). The sample is exposed to linear polarization at a wavelength of 473 nm. A compromise regarding the laser beam intensity is selected, balancing the need to avoid damaging the pattern with the requirement for relatively rapid photostructuration of the dewetted pattern (approximately 20 minutes). The laser beam has a 4 mm diameter at $1/e^2$ with an $M^2$ factor of less than 1.2, a spectral linewidth less than $10^{-5}$, and a power stability of less than 1%, delivering 100 mW of power.

The proposed mechanism for explaining SRG (Surface Relief Grating) formation with a single laser beam within the droplet region is rooted in a reversible trans-cis-trans isomerization, which induces molecular displacements and anisotropic random walks [32]. The self-structuring that ensues typically relies on the laser beam's wavelength, the angle of incidence, and the polarization direction. When the single incident laser beam illuminates the dewetted droplet, an azopolymer grating emerges as a periodic surface modulation with varying amplitudes and pitches. This grating spontaneously forms on the azopolymer film, featuring a spatial wave vector perpendicular to the laser's polarization direction. This pattern is attributed to two counter-propagating transverse-electric (TE) waves at the azopolymer interface [33]. Within the crack domains, a self-organized grating is spatially confined by the substantial and deep grooves of the cracks. The self-organized, photoinduced surface grating within the cracks has a pitch of 950 nm ± 20 nm with an amplitude of 80 nm ± 10 nm (see Fig. 3f). This surface grating aligns with the laser's polarization and is oriented perpendicular to the polarization direction under investigation. The overall pattern exhibits very minimal surface roughness, with the amplitude of the photoinduced grating measuring 80 nm. The depth of the crack structures is approximately 50 nm.

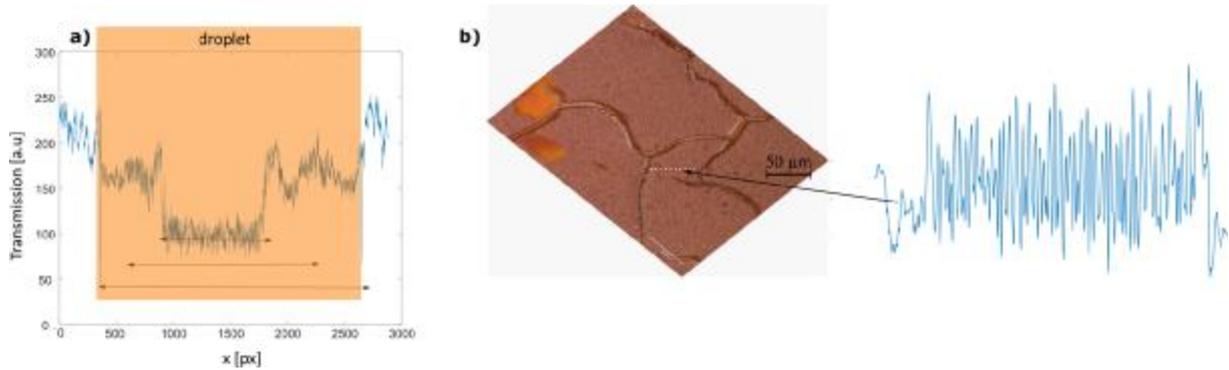

Fig. 4: a) transmission intensity through the mask, b) detail of the topography of the surface grating.

3. Sample illumination experiments

The sample, consisting of a self-photoinduced patterned dewetted droplet on a glass substrate, was utilized as an amplitude mask for beam shaping (see Fig. 4). Figure 4a illustrates the light transmission intensity through the mask. The central section of the mask comprises the surface grating with its topography, as presented in Figure 4b. A collimated laser beam (1 mW DPSS source, Diode Pump Solid State) with a 1 cm diameter and a divergence of less than 1.5 mrad illuminates the mask. Its wavelength, 532 nm, was chosen well outside the material's absorption spectrum to prevent alterations in the self-structured pattern of the mask during the experiment (the transmittance of the azopolymer pattern is nearly 60%). The original beam diameter was reduced using a pupil diaphragm to only illuminate the pattern inscribed in the droplet, while the laser intensity was controlled using a half-wave plate and a polarizing crystal. The experimental setup is outlined in Figure 5.

Upon reaching the dried droplet on the glass substrate, the incoming beam undergoes diffraction. The experiment does not consider any phase modulation. A CCD camera placed at various distances from the masked sample was employed to observe the propagated beam through the diffraction pattern as a function of distance.

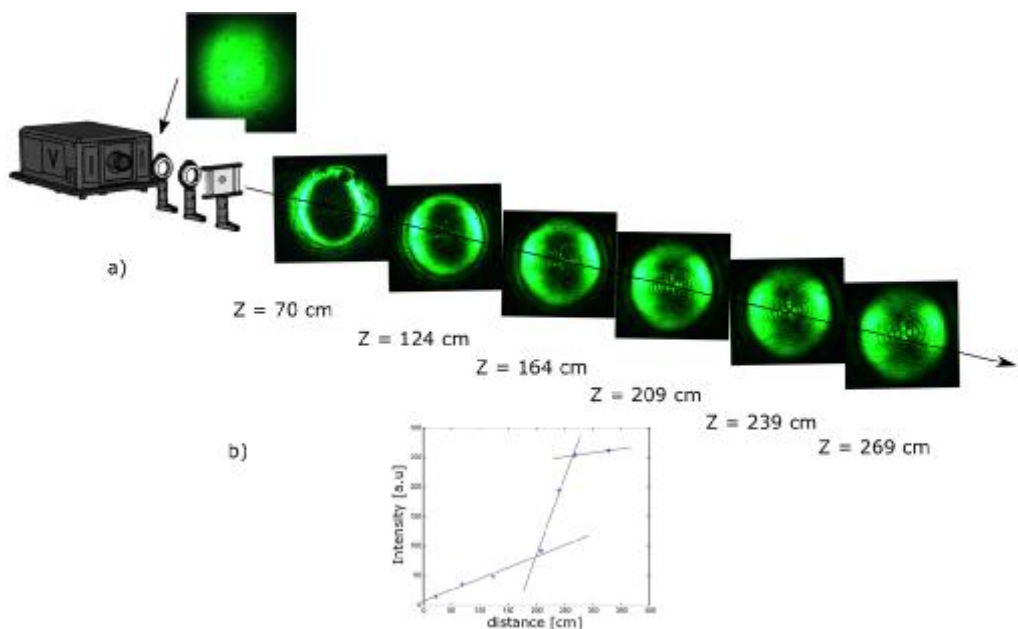

Fig. 5. a) Evolution of the observed laser beam after transmission through the dewetting pattern used as an amplitude mask and as a function of the distance, b) Intensity of the central pattern as a function of the distance.

The diffracted irradiance was recorded in close proximity to the sample (a few centimeters) and in the far field (several meters) along the beam path. Near the diffractive mask, the laser irradiance distribution exhibits a ring shape. As the distance increases and the beam propagates, the light pattern converges toward a final irradiance pattern.

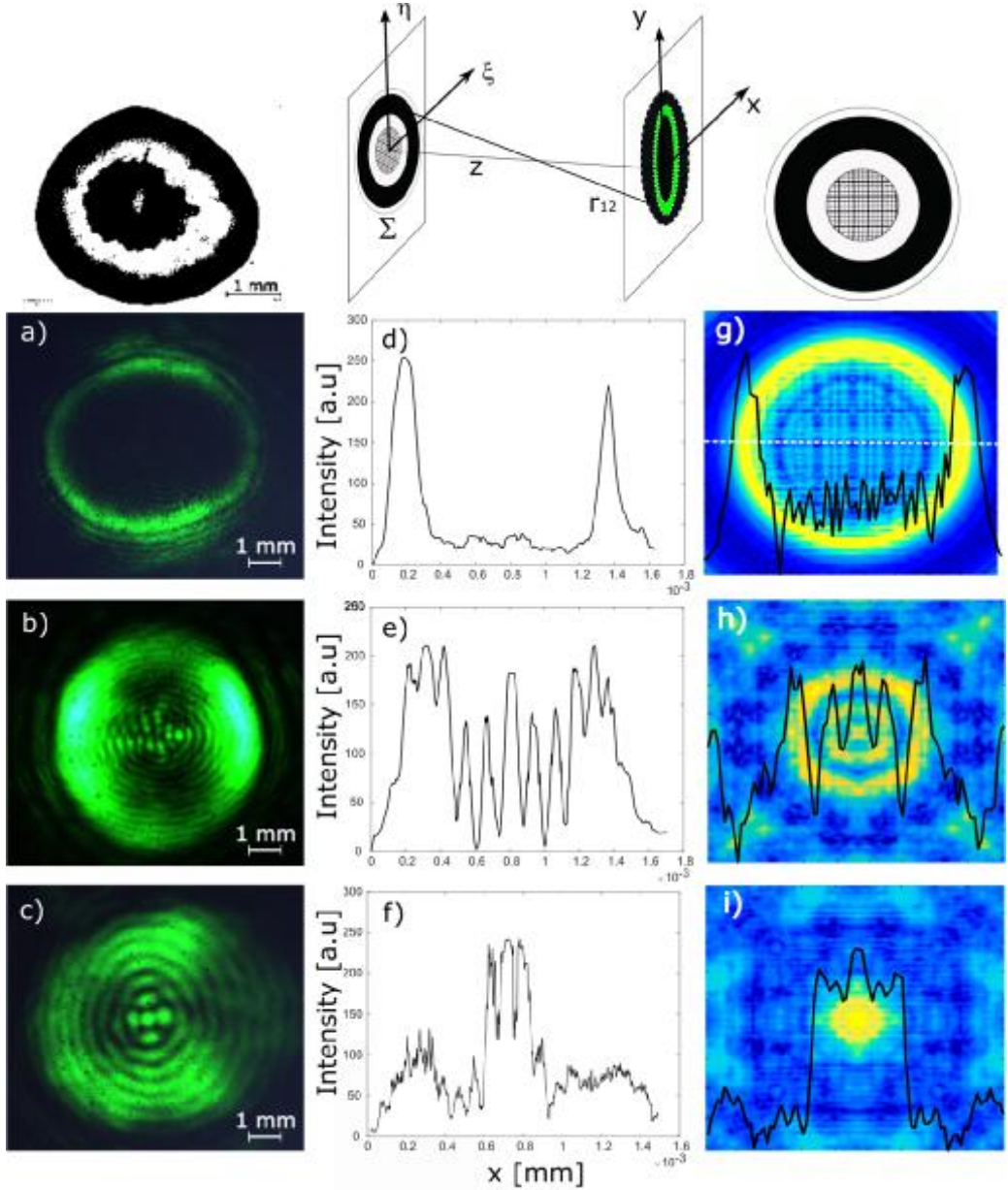

Fig. 6. Comparison of experimental illuminated patterns for different distances with simulations. a) and d) close to the mask (5 cm), b) and e) at a distance of 200 cm, c) and f) at a distance of 440 cm. The observed images are obtained with a beam size reduction of 3.15. For the simulation, the dewetted pattern illuminated with the laser is designed by a simplified mask simulating the different, diffracting contributions after propagation as a function of the same distance of the experiments, g) close to the mask, h) at a distance of 200 cm and i) at 440 cm. In the simulations the white lines represent the plot of the cross-section intensities.

A lens with a magnification of 3.15 is employed to match the CCD sensor's area. To analyze the changes in the pattern along the laser propagation, cross-sections of the diffraction images are compared as a function of the distance (see Fig. 6). When the beam is transmitted through the mask along the z-direction and is close to the mask, the inner part of the ring beam is vacant, as illustrated in Fig. 6. Then, at a greater propagation distance, the main portion of the central beam becomes populated with beamlets, and the central spots begin to grow in size, reaching maximum intensity (Fig. 6f). The light intensity spots in the pattern generated by the diffracted beam significantly increase at a threshold point, approximately at $z_{th}$ = 200 cm ± 15 cm, with small spots appearing inside the rings. This threshold is defined when a noticeable rearrangement of light intensity becomes apparent. Up to the position $z_{th}$, most of the beam energy is concentrated in a ring structure; afterward, the light intensity starts to be redistributed, resembling a more complex distribution of spots in the center (see Fig. 6(c)). The intensity at the center of the ring begins to increase, reaching 50% of the maximum intensity at z = 50 cm. Figures 6(a), 6(b) and 6(c) display the diffraction patterns recorded with the CCD camera. The spots within the ring become visible at a distance in the range of 200 – 220 cm from the sample (Figs. 6(b) and 6(e)).

In the far field, the diffraction pattern becomes more intricate, acquiring additional intensity variations. The contribution of the central part of the mask then becomes dominant. This variability in the light pattern stabilizes after 260 cm (see Fig 6(c) and 6(f)). Consequently, for longer propagation distances, the beam maintains the same profile with a beam divergence of only 80 mrad.

The observed profile of the generated pattern in the far field is a result of the outer ring of the dewetted pattern, which limits the transverse expansion of the transmitted beam and the diffraction generated by the photo-structuration within the dewetting pattern and cracks. After a few centimeters, the diffracted beamlets interfere and self-organize to shape the light through the modified pattern in the far field. The intensities of the rings are uniform, in contrast to a Bessel beam. A central spot is observed with a diameter of 5% of the whole beam diameter, surrounded by four lobes, each with the same intensity and diameter. The light intensity is then divided into equal circular intensity rings. A dewetted pattern without an inscribed SRG does not exhibit such diffracted intensity lobe arrangements; instead, a conventional Bessel beam is observed with the usual rings. The significant formation of crack domains and self-structuration inside them is responsible for the appearance of the diffracted lobes.

To enhance our comprehension of the generated diffracted optical patterns observed after the mask, optical simulations were performed at various propagation distances (refer to Fig. 6). For these simulations, a source plane is employed with coordinate variables ξ and η, and Σ denotes the illuminated aperture. In the source plane, the distribution of the optical field is represented as $U_1(\xi, \eta)$. Following propagation, the observation plane $U_2(x, y)$ is calculated using the first Rayleigh-Sommerfield diffraction solution.

$$U_2(x,y) = \frac{z}{j\lambda} \iint_\Sigma U_1(\xi,\eta) \frac{\exp(jkr_{12})}{r_{12}^2} d\xi d\eta$$

where l is the optical wavelength, k the wavenumber, z the propagation distance and $r_{12}$ the distance between a position on the source plane and a position in the observation plane. The distance $r_{12}$ is:

$$r_{12} = \sqrt{z^2 + (x - \xi)^2 + (y - \eta)^2}$$

If the source and observation areas are defined on parallel planes, in the paraxial optical system, the optical wave through the droplet pattern can be described by the Huygens–Fresnel diffraction integral, which can be written as:

$$U_2(x, y) = \frac{z}{j\lambda} \iint_\Sigma U_1(\xi, \eta) \frac{\exp(jk\sqrt{z^2 + x^2 + y^2})}{z^2 + x^2 + y^2} d\xi d\eta$$

$U_2(x, y)$ can be written with the Fourier convolution theorem:

$$U_2(x, y) = Á^{-1}\{Á\{U_1(x, y)\}Á\{H(f_x, f_y)\}\}$$

where H is the Rayleigh–Sommerfeld transfer function. $U_2$ is calculated with matlab software where $U_1$ is the source aperture model defined in the figure 6 as a representation in term of binary image as similar as possible to the droplet pattern.

The initial diffracting pattern used for computer simulations is an amplitude mask represented with a simple design (see Fig. 6): the dewetted mask from Fig. 4(a). The coffee ring corresponds to the surrounding white annular section of the designed mask, while the grating and crack structures are depicted in the central region with intersecting lines of varying thickness.

In the simulations, we assume an incident plane optical field transmitted through the mask at z = 0, after which the field propagates freely. The propagation was numerically computed using the Fourier convolution theorem method [34] in Matlab. In the near field, the Fresnel approximation is employed, while the Fraunhofer approximation is used for longer propagation distances.

Figures 6(g)–(i) demonstrate that the patterns obtained in the experiment can be effectively simulated and replicated using a simple mask consisting of a ring and a rectangular grating inscribed inside a centered circle. Figure 6(g) displays a central ring, which is also observable in the cross-section of Fig. 6(d); the difference in the intensity of the central ring could be attributed to the absorption of the central part of the experimental pattern.

In Fig. 6(h), the diffracted pattern begins to exhibit structures at the center while annular structures are still present. As the distance of propagation increases, Fig. 6(i) resembles the central spot with the four lobes, corresponding to Fig. 6(c). Other features of the diffracted beam present in the simulation are absent in the experiments. These additional features appear to be superimposed on the rings formed by the experimental diffracting pattern.

Additional simulations have been carried out with the aim of elucidating the contribution of the various structures within the mask. The outer ring in the diffraction profile results from the coffee ring structure, while the four central spots result from the central grating. In other words, the interaction between the outer ring and the central grating contributes to the observed pattern. Various simulations involving different sizes and ratios of the outer ring and the central pattern were conducted to determine the optimal size of the designed mask (Fig. 7), capable of replicating the experimental observations.

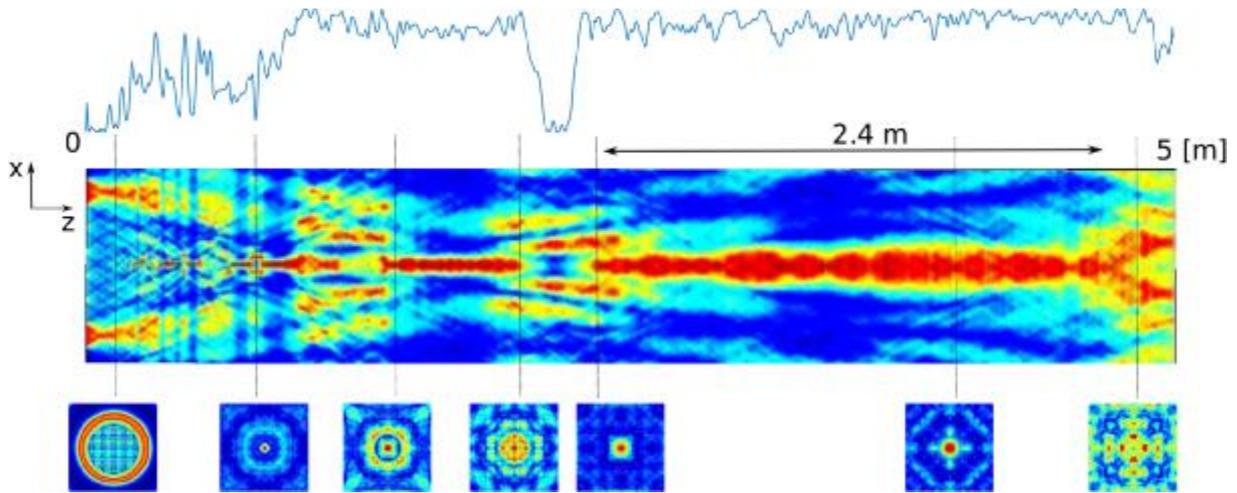

Fig. 7. 2D simulation of the evolution of the laser propagation in xz plane after the transmission of a plane wave through the designed mask representing the evaporated photoinduced pattern. The different steps along the propagation are represented by different cross-sections.

The radius of the propagated beam at various distances is simulated, as illustrated in Fig. 7, outlining a multi-self-focusing trajectory (3 in the propagated beam simulation). It is observed that, as z increases, the beam's radius contracts along a parabolic path. Subsequently, the beam sharply focuses on a specific point, corroborating the experimental results. The autofocusing observed in the experiment aligns well with the numerical simulation. Figure 7 also presents a numerical simulated side-view propagation of the new beam in free space, with different cross-sections depicting the beam's evolution as a function of distance. The distances are selected as close to the mask, at the threshold, and far from the mask, with intermediate positions between the mask and the threshold, as well as between the threshold and the far field. The beam transforms into a shape resembling a bottle with a pillar in the middle, and then it converges into an array of central spots. It is interesting to point out the long distance of propagation of the non-diverging beam (~ 2.4 m). Furthermore, this transition with long non diverging beam is repeated along the propagation as it is observed two non-diverging profile of the beam in the simulated propagation length.

4. Conclusion

A dewetting pattern of an azopolymer droplet is presented with two distinct structures: an outer ring and a central part. During droplet evaporation, two counter-propagating laser beams with a wavelength within the absorption band of the azopolymer work to reduce the coffee-ring effect by accelerating the evaporation rate at the droplet's apex much faster than on other surfaces, thus allowing for the creation of a central part. The strong evaporation rate induced in the central region of the droplet provides the driving force for the Marangoni flow, which, in turn, attenuates the coffee-ring effect. The dewetted droplet consists of two separate regions. The central part has a greater thickness compared to the outer part. Laser illumination of this central section initiates photoinduced self-structuring within the material's crack formation, inducing a surface relief grating. The resulting pattern is then used as an amplitude mask for laser beam shaping. The diffracted pattern observed near the sample displays rings, while in the far field, a laser pattern emerges, featuring rings of equal intensity and central laser spots. Optical simulations were conducted using a simplified design of the experimental amplitude mask, enabling a clear understanding of the observed laser beam pattern and confirming the experimental results. The transmitted beam through this mask is cost-effective, simple to fabricate, and exhibits minimal divergence. This novel laser beam shape is anticipated to reduce beam wandering after transmission in atmospheric turbulence, offering a potential solution for mitigating strong perturbations in turbid media.

Acknowledgement
 D. G. P has a support from Agencia Nacional de Investigación y Desarrollo (ANID), Chile (FONDECYT 1211848; ANILLO ATE220022; ECOS220012); Pontificia Universidad Católica de Valparaíso (PUCV), Chile (Project 123.774/2021). R. B gratefully acknowledges the financial support of the program ECOS-Sud and PHC Polonium.

Data availability.
Data underlying the results presented in this paper are not publicly available at this time but may be obtained from the authors upon reasonable request.

Disclosures
The authors declare no conflicts of interest.